# Symmetries and selection rules in photoelectron chiral dichroism from tailored light

Ofer Neufeld[*]

Technion Israel Institute of Technology, Faculty of Chemistry, Haifa 3200003, Israel.
*Corresponding author E-mail: ofern@technion.ac.il

Photoelectron circular dichroism (PECD) is a method whereby randomly oriented chiral molecules are irradiated by circularly-polarized light, photoionizing electrons, which are measured in a momentum-resolved manner. This scheme permits chiral light-matter interactions within the electric-dipole approximation (avoiding weak magnetic-dipole interactions), yielding huge chiral signals in the form of a forwards-backwards asymmetry in photoemission. Recently, more intricate realizations of PECD have been explored where the circularly-polarized light is replaced with elaborate polarization-tailored light (e.g. bi-chromatic, non-collinear, etc.), some of which do not even require circularly-polarized components, but still lead to massive chiral signals. However, the connection between generalized symmetries and asymmetries of the laser drive and selection rules in PECD have not yet been derived. Here we formulate this connection analytically from group theory, also predicting two previously unknown selection rules for PECD from fields with dynamical inversion and improper-rotational symmetries. We further propose bi-chromatic bi-elliptical fields for breaking symmetries in typical spectra, yielding potentially more information for ultrafast-resolved measurements. We numerically demonstrate all of our results with state-of-the-art *ab-initio* simulations in the archetypal chiral molecule, Bromochlorofluoromethane, providing predictions for experiments. Our work presents a roadmap for analyzing PECD from tailored light and resolved a long-standing issue in the field. It should motivate theoretical and experimental investigations in novel PECD set-ups.

## I. INTRODUCTION

Chirality is an inherent asymmetry of objects that cannot be superposed onto their mirror image[1]. It arises naturally in the universe at all length scales, and is especially important in material science[2, 3], chemistry[4], and biology[5, 6]. In chemistry, chirality manifests in chiral molecules, which form inverted molecule pairs called enantiomers. Enantiomers have identical physical and chemical properties (e.g. energy landscape, reactivity, absorption, etc.), unless when interacting with another chiral object such as another chiral molecule or circularly-polarized light (CPL). Historically, interactions with CPL were the main approach to detect and quantify chirality[7]. For instance, chiral signals are found in polarization rotation, or circular dichroism (CD) in absorption/transmission. However, such methods are limited in their scope since chiral signals are usually very small (order of ~0.001-0.1%). Physically, this is a result of CD requiring interactions with magnetic-dipoles, which are generically weak. This restricts typical CD methods in analyzing various chiral systems of interest, from gas phase chirality, ultrafast chirality[8–12], parity violations[13, 14], multi-chiral center molecules for emerging drugs[15], and more. Consequently, in the last decades many techniques overcoming this so-called magnetic-dipole conundrum were developed[8, 16], providing substantially larger chiral signals. These approaches include, among others, wave mixing schemes[17–19], enantiospecific state transfers[20–22], non-linear optical emission[12, 23, 32, 24–31], Coulomb explosion[33, 34], and using polarization and/or spatially tailored light in various forms[23, 24, 32, 35–37]. This paper deals with the method of photoelectron CD (PECD), which was shown to provide robust chiral signals as large as ~30% in the gas phase[38–43] and ultrafast temporal resolution[10, 44, 45].

In PECD, CPL irradiates randomly oriented chiral media, and the resulting photoelectron emission is measured in an angle- and energy-resolved manner, providing momentum-resolved ionization yields, $P_{R/S}(\mathbf{k})$, where R/S signifies the molecular handedness, and $\mathbf{k}$ is the photoelectron momentum. By analyzing the normalized differential spectra upon flipping the CPL helicity (or equivalently, the medium's handedness, R→S), defined as: $PECD(\mathbf{k}) = 2\frac{P_R(\mathbf{k}) - P_S(\mathbf{k})}{P_R(\mathbf{k}) + P_S(\mathbf{k})}$, the medium's handedness and other chiral attributes can be extracted. PECD's most prominent signature is a forwards-backwards asymmetry, meaning electrons in chiral molecules tend to ionize preferentially towards (or against) the axis of the CPL beam's propagation (denoted as the *z*-axis here throughout). Upon switching handedness that preferential ionization inverts sign.

In practice, this leads to a symmetry of the PECD spectra (a selection rule), whereby $PECD(k_x, k_y, k_z) = -PECD(k_x, k_y, -k_z)$, making it an odd function of $k_z$. There are other characteristic symmetries, such as an up-down symmetry $(PECD(k_x, k_y, k_z) = PECD(k_x, -k_y, k_z)$ and left-right symmetry $PECD(k_x, k_y, k_z) = PECD(-k_x, k_y, k_z))$, which depend on the exact driving conditions.

Recently, PECD has been predicted(46) and measured(47, 48) in various configurations involving tailored light bi-chromatic driving instead of typical CPL beams. Here multiple carrier waves can be combined to tailor the optical field's time-dependent polarization and instantaneous chirality(49). It was shown that with $\omega$-$2\omega$ beams that are individually linearly-polarized and co-linearly propagating, chiral signals on order of few percent can still be obtained(47). In that case the typical forwards-backwards asymmetry of PECD is upheld, but up-down symmetry is broken and is replaced with up-down asymmetry(46, 47). Since the $\omega$-$2\omega$ beams are not circular, the standard notion of circular dichroism no longer holds. Instead, the equivalent of inverting the medium's handedness (R→S) is acting with a mirror operation on the driving beam's polarization, which requires changing relative phases and potentially other spectral attributes. Hence, we refer from this point on only to chiral dichroism (CD), using the acronyms interchangeably. More recently, non-colinear $\omega$-$2\omega$ beams were developed that possess local-chirality(23, 50, 51), i.e. light that breaks all mirror symmetries within its three-dimensional time-dependent polarization, $\mathbf{E}(t)$, within the electric-dipole approximation. Locally-chiral light should be highly efficient for chiroptical spectroscopy, and has been predicted to lift the typical forwards-backwards asymmetry of PECD, yielding chiral signals also in above threshold ionization (ATI) spectra(52). What remains unclear in this emerging field, which is expected to grow further due to advancements in tailored light(16), is how generalized symmetries of complex laser fields map onto PECD selection rules. Indeed, selection rules theories were developed for nonlinear optical processes such as high harmonic generation(53–55), photocurrent generation(56, 57), multi-photon absorption(58), and ultrafast magnetism(59), but not for photoionization.

Here we bridge this gap and develop a symmetry theory for PECD, presenting a general formula for selection rules of spectra that arise from dynamical symmetries in the driving laser field. We further derive two previously unknown selection rules for cases where the driving beam exhibits improper rotational or inversion symmetry, as well as cases where typical forwards-backwards and up-down symmetries/asymmetries are fully broken, yielding more information in spectra. We show that for dynamical symmetries, selection rules in PECD require several driving laser cycles to develop, similar to other nonlinear optical phenomena(60, 61). All of our results are supported by *ab-initio* time-dependent density functional theory (TDDFT) calculations of PECD from the archetypal chiral molecule, Bromochlorofluoromethane (CBrClFH). Our work paves the way to novel realizations of PECD with intricate light forms for tunable photoelectron spectra and enhanced ultrafast chiral spectroscopy.

## II. METHOLOGY

We begin by outlining our employed theoretical approach. Numerically, PECD is calculated directly with *ab-initio* TDDFT(62) simulations using Octopus code(63–65) with a methodology similar to that in ref. (52). In brief, the interaction of the CBrClFH molecule with the driving laser field, $\mathbf{E}(t)$, is described in the dipole approximation and length gauge by the Kohn-Sham (KS) equations of motion:

$$i\partial_t |\varphi_n^{KS}(t)\rangle = \left(-\frac{1}{2}\nabla^2 + v_{KS}(\mathbf{r}, t) - \mathbf{E}_\Omega(t) \cdot \mathbf{r}\right) |\varphi_n^{KS}(t)\rangle \quad (1)$$

where $|\varphi_n^{KS}(t)\rangle$ is the *n*'th time-dependent KS orbital, $v_{KS}(\mathbf{r}, t)$ is the time-dependent KS potential:

$$v_{KS}(\mathbf{r}, t) = V_{ion} + \int d^3r' \frac{\rho(\mathbf{r}', t)}{|\mathbf{r} - \mathbf{r}'|} + v_{XC}[\rho(\mathbf{r}, t)] \quad (2)$$

Here $V_{ion}$ is an electrostatic interaction of electrons with nuclei and other core electrons (using pseudopotentials(66), and assuming static ions), $\rho(\mathbf{r}, t)$ is the electronic density, and $v_{XC}$ is the XC functional which is taken in the adiabatic local density approximation with an added average density self-interaction correction(67). By solving these equations we obtain the time-dependent KS states, from which photoelectron emission spectrum, $P_{R/S}(\mathbf{k})$, is calculated through the t-surff method(68, 69) implemented in Octopus(70–72).

In eq. (1) $\mathbf{E}_\Omega(t)$ is the driving field oriented along the solid angle $\Omega$ of the molecule, and the resulting PECD spectra is obtained by integrating over $\Omega$ with three Euler angles to obtain orientation-averaged spectra, and subtracting photoemission from R/S enantiomers. The equivalent of velocity map imaging (VMI) spectra are obtained along the three main planes (*xy*, *yz*, *xz*) after integrating photoemission in the transverse axis. That is, PECD in the *xy* plane is obtained by integrating out the $k_z$ component, and similarly for other axes. Most importantly, $\mathbf{E}(t)$ can take any configuration, where we explore below several field geometries with ω-2ω fields, ω-3ω fields, and tri-chromatic ω-2ω-3ω fields, either in collinear or non-collinear configurations. All other technical details of the numerical procedures are delegated to the supplementary material (SM).

### III. PECD SELECTION RULES
#### A. EVEN PARITY IN PECD FROM ROTATIONS

We now analytically explore the connection between PECD and symmetries of the driving electromagnetic field. Within the dipole approximation that is valid for PECD, all symmetries of electromagnetic fields can be cataloged using Floquet group theory(53). There are two relevant classes of symmetries: (i) Static symmetries, where the field is invariant under a point group operation acting on its polarization space. For freely propagating laser beams the only available static symmetry is a mirror plane. For example, CPL propagating along the *z* axis exhibits a trivial mirror symmetry where $\mathbf{E}(t)$ is invariant under $E_z(t) \rightarrow -E_z(t)$ (because $E_z(t) = 0$). In group theory language, this can be written as $\mathbf{E}(t) = \sigma_{yz} \cdot \mathbf{E}(t)$, and we denote $\sigma_{yz}$ as a symmetry for $\mathbf{E}(t)$ and a member in its Floquet group. Linearly-polarized light exhibits $\sigma_{yz}$ symmetry as well, on top of an infinite number of other static mirror planes in its symmetry group, much like in the $D_{\infty h}$ point group for diatomic molecules. (ii) Dynamical symmetries(53, 73–75), where the field is invariant under combined temporal (e.g. time-translation) and point group operations (e.g. rotations, improper rotations, etc.) acting on the field's polarization space. Here the wealth of potential dynamical symmetries is enormous, and many types of symmetric fields have been generated and employed for nonlinear optics. For instance, an orthogonally polarized ω-2ω field(76–79), $\mathbf{E}(t) = E_0[\cos(\omega t + \phi)\hat{\mathbf{x}} + \Delta \sin(2\omega t)\hat{\mathbf{y}}]$ (where $\phi$ is a relative phase and $\Delta$ is the amplitude ratio between the carrier components), exhibits a dynamical mirror plane $\sigma_{yz}$ such that the field is invariant under $t \rightarrow t + T/2$ (denoted $\tau_2$), $E_x \rightarrow -E_x$, where $T = 2\pi/\omega$ is the temporal period of the ω beam. In other words, for this driving field configuration we have $\mathbf{E}(t) = \sigma_{yz} \cdot \mathbf{E}(t + T/2) = \tau_2 \sigma_{yz} \cdot \mathbf{E}(t)$, with $\sigma_{yz}$ acting on the field polarization components and $\tau_2$ on its temporal dependence.

At this point we wish to ask how, and if at all, symmetries of $\mathbf{E}(t)$ map onto selection rules in PECD, which requires also the medium's point of view. For PECD the medium is an ensemble of randomly oriented chiral molecules that is fully symmetric under O(3) – it is invariant under all possible rotational point group operators, but inherently does not respect any mirror, inversion, or improper rotational elements. Overall, the full light-matter Hamiltonian describing the interaction of randomly oriented chiral molecules with an electric field can therefore only ever be invariant under dynamical symmetries involving pure rotations. These form the first class of PECD selection rules, which we denote as PECD 'symmetries': If $\mathbf{E}(t)$ exhibits a rotational dynamical symmetry (such as the well-known three-fold dynamical symmetry of the $\omega$-$2\omega$ counter-rotating bi-circular field(80–82)), it gets mapped onto photoelectron spectra (in both R and S molecular ensembles individually) that becomes rotationally invariant in the plane transverse to the rotational axis. Basically, if $R_n$ (a rotation by $2\pi/n$ along any axis) coupled to a time translation is a symmetry element of the field (e.g. $\tau_n$), then the resulting PECD spectra is also invariant under the spatial part of this dynamical symmetry, $R_n$.

An example of this selection rule is the well-known up/down and left/right symmetries of standard PECD driven by CPL – CPL exhibits a continuous rotational DS. After integrating transverse components in VMI measurements to generate PECD in the *zy* and *yz* planes, a pure up-down or left-right symmetry is recovered, because the PES is invariant under 180 rotations in both enantiomers individually. Trivially, this can be understood since the CPL photoionizes an equal amount of electrons both up and down, and left and right, when considering the time-integrated measurement. However, the group theory perspective provides a mathematical basis that can be more easily employed in complex cases.

B.     ODD PARITY IN PECD FROM IMPROPER ROTATIONS

On the other hand, PECD does not arise solely from interactions with just one molecular handedness, but as a subtraction of photoemission from R/S-handed molecular ensembles. Since both enantiomeric ensembles are intrinsically connected *via* any improper rotational operation (here it is worth noting that improper rotations include also inversion symmetry, which is an improper rotation of order $n$=2, and mirror planes, which are improper rotations of order $n$=1), any improper-rotational dynamical symmetry of $\mathbf{E}(t)$ still maps onto a PECD selection rule. This is despite the fact that the Hamiltonian of a given handedness enantiomeric ensemble interacting with light is never invariant under dynamical improper rotations. Let us explain how this comes about mathematically: Improper rotations, $S_n$, can always be separated into sequential point group operations of rotation, $R_n$, times a mirror operation in the transverse plane $\sigma_h$. $R_n$ leads to a 'symmetry' in PECD through the first class of selection rules described above. $\sigma_h$ effectively acts as a molecular handedness replacement $R \to S$, since both enantiomers are connected to one another through it, which in the PECD expression gives a minus sign due to the subtraction of $R$ and $S$ terms. Practically, this means that the mirror symmetry part of the improper rotational symmetry leads to an odd function in PECD along the axis transverse to $\sigma_h$. Overall, this forms the second class of PECD selection rules, which we denote as PECD 'asymmetries', i.e. odd parity behaviors in $PECD(\mathbf{k})$.

The most well-known example is the typical forwards-backwards asymmetry in PECD – in our group theory framework it arises since CPL exhibits a static mirror plane $\sigma_{xy}$. Then the operation $k_z \to -k_z$ connected with the mirror plane in the PECD spectra maps onto a minus sign selection rule, meaning that $PECD(k_x, k_y, k_z) = -PECD(k_x, k_y, -k_z)$ as discussed above. Notably, $\sigma_{xy}$ arises for any collinearly-propagating laser beam configuration, meaning that forwards-backwards asymmetry is a generic feature in PECD unless the driving field is polarized in the full three dimensions (with a nonzero $E_z$ component). Alternatively, if the light does not exhibit any such mirror symmetry as in locally-chiral light(23, 50), then the resulting PECD spectrum is not forwards-backwards asymmetric(52).

Another previously measured example is PECD driven by the orthogonally polarized $\omega$-$2\omega$ field discussed above, which exhibits a dynamical $\tau_2 \sigma_{yz}$ symmetry, on top of the static $\sigma_{xy}$. Indeed, in refs. (46, 47) it was shown that the static $\sigma_{xy}$ leads to the typical forwards-backwards asymmetry, but also to an up-down asymmetry: $PECD(k_x, k_y, k_z) = -PECD(-k_x, k_y, k_z)$. This was explained ad-hoc based on instantaneous optical chirality considerations(49), which in this particular cases flips sign between the half-cycles where $\mathbf{E}(t)$ points up or down. However, from the symmetry picture we can understand the feature arises more generally due to the additional dynamical $\sigma_{yz}$ symmetry in the tailored laser field.

To summarize this part, our framework maps any dynamical symmetry in $\mathbf{E}(t)$ that is static, or involves time-translational elements, to a selection rule in PECD: rotational parts of the symmetry operator under which $\mathbf{E}(t)$ is invariant map onto the photoemission and PECD spectra itself, making it invariant under the same rotational operator, and mirror operations map as a minus sign in the PECD spectra transverse to the mirror plane. Then, composite dynamical symmetries can be decomposed to rotations and mirror planes to uncover underlying selection rule. This approach reproduces all known PECD symmetries and asymmetries, and in the next section we will show that it also predicts selection rules that were previously unknown.

Lastly, we highlight that we excluded from this discussion dynamical symmetries of $\mathbf{E}(t)$ that involve time-reversal elements, which for instance have been shown to lead to selection rules in photocurrent generation(56, 57). However, time-reversal elements cannot lead to PECD selection rules. This is because chiral signals are inherently time-even pseudoscalars(8), and in the case of PECD the photoionization process itself is not invariant under time reversal (since photoemission is a different process from recombination with difference cross-sections).

## IV. SYMMETRY AND ASYMMETRY IN PECD

We now present an *ab-initio* investigation of the above analytical theory. We analyze two new PECD configurations where the driving field exhibits dynamical inversion symmetry, and dynamical improper rotation symmetry, but does not include any other symmetry elements in its Floquet group. First, we consider a non-collinear $\omega$-$3\omega$ beam configuration where the $\omega$ and $3\omega$ beams are both circularly-polarized and co-rotating, but have a small opening angle in between their propagation axes (which is slightly offset from the z-axis, see illustration in Fig. 1(a)). In this configuration $\mathbf{E}(t)$ exhibits a dynamical inversion symmetry $\mathbf{E}(t) = -\mathbf{E}(t + T/2)$, which can be expressed as $i\hat{\tau}_2$, with $i$ an inversion operator. However, for the sake of the selection rule derivation we note that inversion can be decomposed into a two-fold rotation times a transverse mirror operation: $i = R_2 \sigma_h$. Note that the rotational axis can be any spatial axis in this decomposition, as inversion is a symmetry with respect to a point. Since $\mathbf{E}(t)$ in this configuration is no longer polarized in a plane, static $\sigma_{xy}$ symmetry that usually gives rise to forwards-backwards asymmetry in PECD should vanish, which presents another test for our analytic theory.

Figures 1(b-d) show *ab-initio* calculated PECD spectra from CBrClFH in the various planes after averaging over the transverse axis (mimicking VMI measurements). In both *xz* and *yz* planes no forwards-backwards asymmetry appears, validating the prediction from the analytical group theory. This result is along similar lines of broken forwards-backwards asymmetry from locally-chiral light, which is driven in the same configuration but with $\omega$-$2\omega$ beams instead(52). Notably, for the locally-chiral $\omega$-$2\omega$ case the driving tailored field does not exhibit any symmetry relations at all, which has been shown also to lead to chiral signals in ATI and total ionization yields. In that case no selection rules appear in PECD spectra. Here on the other hand, the spectra still show a degeneracy arising from the underlying inversion symmetry: Along all planes the PECD is odd under a two-fold rotation. For instance, in the *xy* plane this amounts to: $PECD(k_x, k_y, k_z) = -PECD(-k_x, -k_y, k_z) = -R_2 \cdot PECD(k_x, k_y, k_z)$, which is highlighted by guiding arrows in Fig. 1(b). Similar relations hold in the other planes, as predicted by the analytical theory (since inversion symmetry does not refer to a specific plane). This numerically validates the prediction for a new selection rule in PECD driven by non-collinear tailored light.

Let us highlight a noteworthy feature of the associated selection rule – it is not numerically exact, but approximate. This can be seen as small deviations appearing in Fig. 1 along regions connected by guiding arrows, and is markedly different from the typical situation of forwards-backwards asymmetry in PECD. The reason is that usual forwards-backwards asymmetry in PECD arises from a static symmetry, which is upheld exactly in the Hamiltonian. The dynamical inversion symmetry on the other hand arises in connection with time-translation operators that are not exact due to the driving field being pulsed and having a temporal envelope. As the field duration is taken longer and longer, approaching steady-state conditions, deviations from the selection rules should diminish, as has been shown in other dynamical symmetry related selection rules(53, 60, 61). We explore this relation by calculating the total ionization CD ($I_{CD} = \int PECD(\mathbf{k})d\mathbf{k}$) in this configuration, which formally should vanish since the driving field exhibits dynamical inversion symmetry. That is, after integrating the photoemission angularly, both enantiomers should yield the same ionization rate such that $I_{CD} = 0$. Figure 2 presents the calculated ionization CD vs. the duration of the $\omega$-$3\omega$ driving pulse, showing that indeed the dynamical inversion symmetry is very closely upheld, but not exactly. $I_{CD}$ can be as large as ~0.015% for short few-cycle driving (with full width half max of 2 optical cycles), and substantially diminish to ~0.001% for sufficiently long-duration driving. Overall, our results establish an adiabatic relation for upholding dynamical symmetry selection rules in PECD, and uncover that total ionization CD can be obtained with few cycle non-collinear driving, which is potentially applicable for optical enantio-purification.

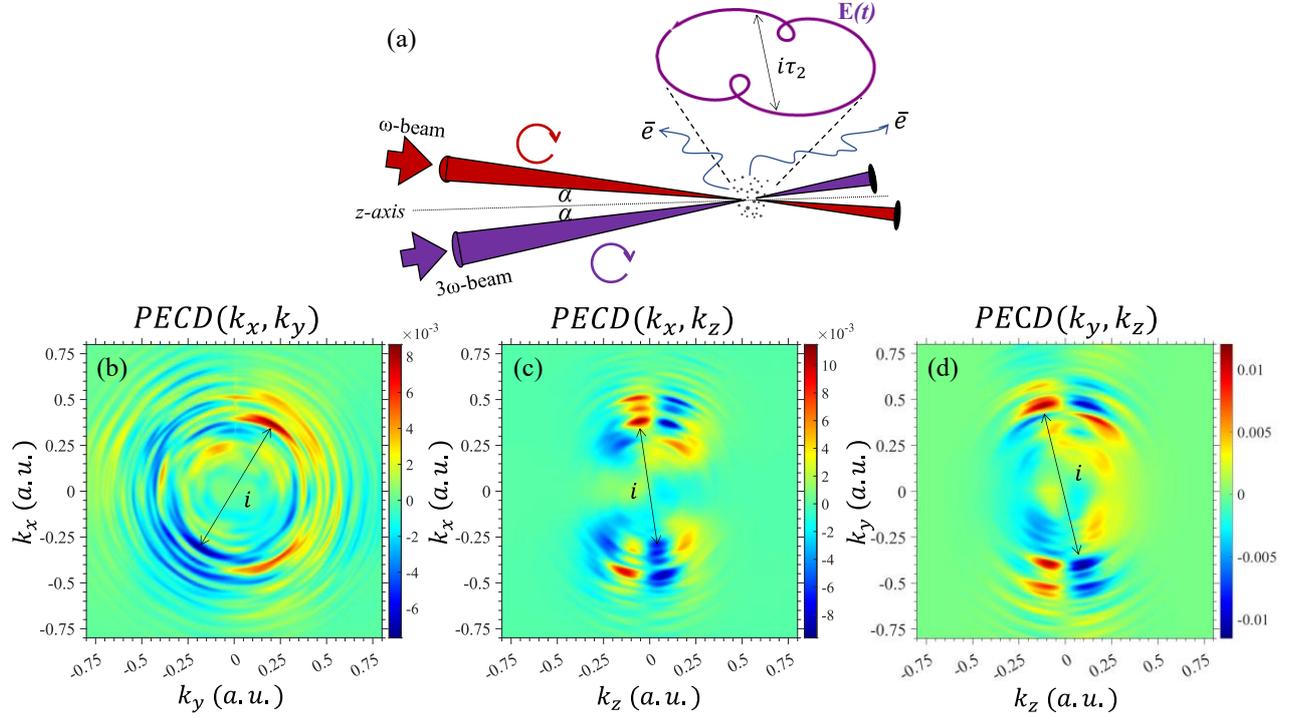

FIG. 1. PECD from CBrClFH driven by ω-3ω non-collinear tailored light exhibiting dynamical inversion symmetry. (a) Illustration of method and beam set-up. (b-d) PECD spectra along three main planes $xy$, $xz$, $yz$. Arrows indicate regions connected by the dynamical symmetry-induced selection rules (whereby PECD should be odd under $R_2$ rotations, see discussion in text). For simulations we have taken the laser power of the ω beam to be $5\times10^{13}$ W/cm$^2$, the beam power ratios to be 1:4, ω corresponds to a wavelength of 800nm, the beam polarizations are co-rotating and circular, the total beam opening angle is 10 degrees, and the pulse duration is taken to be 8 cycles of the fundamental period $T$.

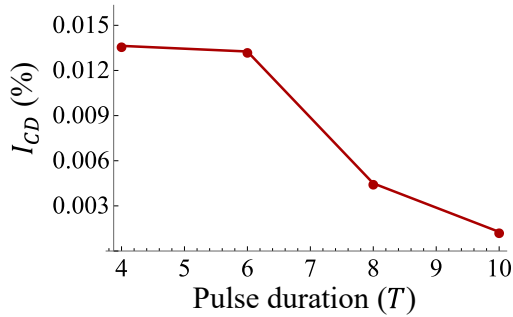

FIG. 2. Total ionization chiral dichroism, $I_{CD}$, calculated in the same conditions as in Fig. 1 but *vs.* driving pulse duration.

In the next step, we explore a more complex crossed-beam set-up(83), in which one can obtain tailored-light exhibiting dynamical improper rotational symmetry of order $n = 4$. For this we employ three carrier waves in the driving laser: an ω beam propagating along the $z$-axis which is circularly-polarized, a 3ω beam which is co-propagating and counter-circularly polarized, and a 2ω beam which propagates transversely along the $x$-axis, and is linearly-polarized along the $z$-axis (see illustration in Fig. 3(a)). This creates an overall tailored driving field that spans all three polarization directions and invariant under $\tau_4 S_4$, with the improper rotational axis being the $z$-axis. Following the analytical prescription above, the resulting PECD selection rules should lead to PECD spectra that is odd (i.e. asymmetric) under a four-fold rotation in the $xy$ plane. Since $R_2$ is also an inherent symmetry in this system, the spectra in the $xz$ and $yz$ planes should also be up-down symmetric. Figures 3(b-d) present *ab-initio* calculated PECD from CBrClFH, which conforms to all of these selection rules. Notably, the spectra in Fig. 3 are not forwards-backwards asymmetric, as that typical selection rule does not exist in these driving conditions.

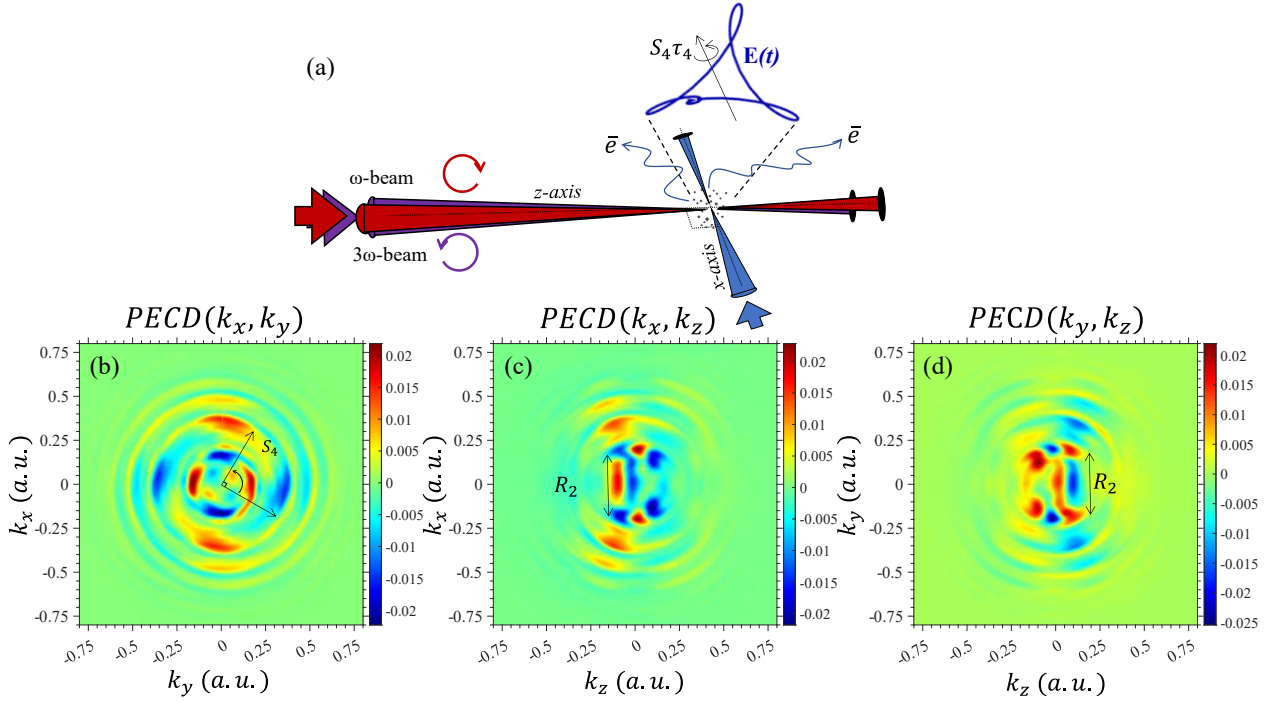

FIG. 3. PECD from CBrClFH driven by ω-2ω-3ω cross-beam tailored light exhibiting dynamical improper rotational symmetry of order *n=4*. (a) Illustration of method and beam set-up. (b-d) PECD spectra along three main planes *xy*, *xz*, *yz*. Arrows indicate regions connected by the dynamical symmetry-induced selection rules (whereby PECD should be odd under $R_4$ rotations in the *xy* plane, as well as even under $R_2$ in *xz* and *yz* planes, see discussion in text). For simulations we have taken the laser power of the ω beam to be $5\times10^{13}$ W/cm$^2$, the beam power ratios to be 1:2:4, ω corresponds to a wavelength of 800nm, the beam polarizations are co-circular for the ω and 3ω beams, and linear for the 2ω beam along the *z*-axis, and the pulse duration is taken to be 8 cycles of the fundamental period *T*.

Lastly, we test the analytical theory by exploring co-linear ω-2ω driven PECD where both beams have a generic elliptical polarization and arbitrary relative angle. Specifically, Figs. 4(b-c) presents PECD from an ω beam that is elliptically polarized with an ellipticity 0.2, superimposed with a 2ω beam with an ellipticity 0.4, and where the elliptical major axes between beams have a relative angle of 45 degrees. This creates a tailored field that respects $\sigma_{xy}$ as in typical PECD (yielding forwards-backwards asymmetry), but which generally does not exhibit other symmetry or dynamical symmetry relations. The resulting PECD in the *xz* and *yz* planes does not exhibit any selection rules besides forwards-backwards asymmetry (specifically, no up/down, left/right, $R_2$, odd or even behavior, etc.). Such asymmetric tailored-light configurations can yield more spectral information in PECD that can potentially be employed for tracking ultrafast chiral signals (whereas in typical cases different regions in spectra carry the same information as they are limited by the PECD selection rules). This is especially promising if the chiral signals are tracked with respect to some of the driving field parameters, e.g. ellipticities, relative phases, relative angles, etc., creating multi-dimensional spectroscopic configurations. Notably, we show that such symmetry breaking can be employed already in a colinear geometry that is straightforward (unlike non-colinear set-ups that can also break symmetries in PECD, but are more challenging and prone to phase matching issues).

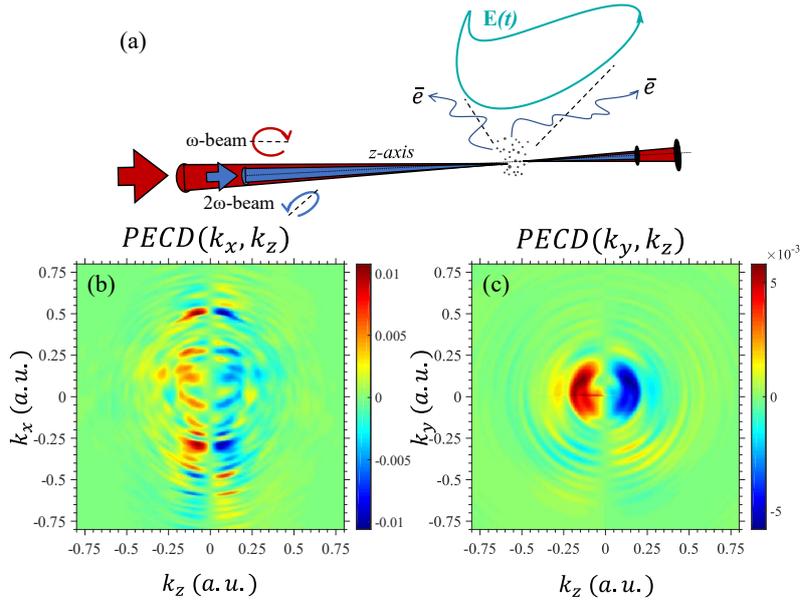

FIG. 4. PECD from CBrClFH driven by ω-2ω co-linear tailored light exhibiting no dynamical symmetry relations besides the typical σ$_{xy}$ that leads to forwards-backwards asymmetry. (a) Illustration of beam set-up and asymmetry. (b-c) PECD spectra along two main planes *xz*, *yz* (here the PECD in the *xy* plane identically vanishes due to the co-linear configuration). For simulations we have taken the laser power of the ω beam to be $5\times10^{13}$ W/cm$^2$, the beam power ratios to be 1:4, ω corresponds to a wavelength of 800nm, the beam polarizations are co-rotating for the ω and 3ω beams with ellipticities 0.4 and 0.2 respectively, the relative angle between the beams major elliptical axis is 45 degrees, and the pulse duration is taken to be 8 cycles of the fundamental period *T*.

## V.  SUMMARY

To summarize, we developed an analytic group theory for symmetries and selection rules in photoelectron chiral dichroism, resolving an outstanding issue in the field of chirality. We further explored PECD driven by poly-chromatic tailored light set-ups using *ab-initio* simulations in the model chiral molecule Bromochlorofluoromethane. Our simulations validated the analytic theory, and further uncovered novel laser set-ups with previously unknown PECD selection rules: (i) a dynamical inversion symmetry leading to two-fold rotational asymmetry in PECD, and (ii) a dynamical improper rotation leading to a four-fold rotational asymmetry in PECD. We showed that in the dynamical symmetry case, selection rules are only upheld approximately due to the laser field temporal envelope, and improve as the driving pulse duration is increased following an adiabatic theorem. We also investigated symmetry-broken conditions in bi-elliptical colinear driving that yield more information in PECD spectra, which could be potentially useful for enhanced ultrafast spectroscopy. Our work presents a roadmap for tailored-light driven PECD that should motivate theoretical and experimental investigations and connect the field of PECD with other nonlinear optical fields already benefiting from symmetry-induced selection rules and symmetry breaking spectroscopies.

## ACKNOWLEDNGEMETS

ON gratefully acknowledges the scientific support of Prof. Dr. Angel Rubio.

# Supplementary material for: Symmetries and selection rules in photoelectron chiral dichroism from tailored light

Ofer Neufeld

Technion Israel Institute of Technology, Faculty of Chemistry, Haifa 3200003, Israel.

We discuss here additional numerical details of simulations. All DFT simulations were performed using Octopus code(63–65). The Kohn Sham (KS) equations were discretized on a Cartesian grid with spherical boundaries of radius 45 Bohr, where the molecule center of mass was centered at the origin. The molecular geometry was taken at the experimental configuration. Calculations were performed using the local density approximation (LDA) with an added average density self-interaction correction (SIC)(67). The frozen core approximation was used for inner core orbitals, which were treated with appropriate norm-conserving pseudopotentials(66). The grid spacing was converged to 0.39 Bohr.

For time-dependent calculations we employed a time step $\Delta t$=0.18 a.u., and added an imaginary absorbing potential of width 15 Bohr. The initial state was taken to be the system's ground-state. The full photoelectron spectrum from each driving laser orientation was calculated using the t-surff method(68, 69), implemented within Octopus code(70–72). A spherical surface where flux was calculated was positioned at $r$=30 Bohr, and integration was performed with a maximal angular momentum index for spherical harmonics of 80, angular grids were spanned with spacing 1°, $k$-grids were spanned with a spacing of $\Delta k$=0.01 a.u. and up to a maximal energy of 38.75 eV. The orientation averaged PES was calculated by trapezoidal integration similar to the method in ref. (52). Photoelectron chiral dichroism spectra were obtained directly by subtracting the photoemission calculated from mirror image enantiomers. Integration over Cartesian axes and angular grids was performed using Simpson integration. The total ionization rate was calculated directly from the electron density rather than integration over the ATI spectra, since this approach has improved accuracy.

The electric field employed in all cases had the temporal envelope of the following 'super-sine' form(84):

$$f(t) = \left(sin\left(\pi \frac{t}{T_p}\right)\right)^{\left(\frac{\left|\pi\left(\frac{t}{T_p}-\frac{1}{2}\right)\right|}{w}\right)} \tag{S3}$$

where $w$=0.75, and $T_p$ is the duration of the laser pulse which was taken as stated in the main text. This temporal envelope mimics a super gaussian and has a full width half max of $T_p$/2. The field polarization components and other parameters are specified in all cases in the figure captions.

**REFERENCES**

1.  E. M. Carreira, H. Yamamoto, *Comprehensive Chirality* (Elsevier Ltd., 2012).
2.  P. Hosur, S. Ryu, A. Vishwanath, Chiral topological insulators, superconductors, and other competing orders in three dimensions. *Phys. Rev. B* **81**, 045120 (2010).
3.  S.-T. Wang, D.-L. Deng, L.-M. Duan, Probe of Three-Dimensional Chiral Topological Insulators in an Optical Lattice. *Phys. Rev. Lett.* **113**, 033002 (2014).
4.  J. Meyer-Ilse, D. Akimov, B. Dietzek, Recent advances in ultrafast time-resolved chirality measurements: perspective and outlook. *Laser Photon. Rev.* **7**, 495–505 (2013).
5.  I. K. Reddy, R. Mehvar, *Chirality in Drug Design and Development* (CRC Press, 2004).
6.  A. Guerrero-Martínez, J. L. Alonso-Gómez, B. Auguié, M. M. Cid, L. M. Liz-Marzán, From individual to collective chirality in metal nanoparticles. *Nano Today* **6**, 381–400 (2011).
7.  N. Berova, P. L. Polavarapu, K. Nakanishi, R. W. Woody, *Comprehensive Chiroptical Spectroscopy* (Wiley, 2013).
8.  D. Ayuso, A. F. Ordonez, O. Smirnova, Ultrafast chirality: the road to efficient chiral measurements. *Phys. Chem. Chem. Phys.* **24**, 26962–26991 (2022).
9.  D. Baykusheva, *et al.*, Real-time probing of chirality during a chemical reaction. *Proc. Natl. Acad. Sci.* **116**, 23923–23929 (2019).
10. V. Svoboda, *et al.*, Femtosecond photoelectron circular dichroism of chemical reactions. *Sci. Adv.* **8**, eabq2811 (2023).

11. S. Beaulieu, *et al.*, Attosecond-resolved photoionization of chiral molecules. *Science* **358**, 1288–1294 (2017).
12. R. Cireasa, *et al.*, Probing molecular chirality on a sub-femtosecond timescale. *Nat. Phys.* **11**, 654–658 (2015).
13. M. Quack, J. Stohner, Influence of Parity Violating Weak Nuclear Potentials on Vibrational and Rotational Frequencies in Chiral Molecules. *Phys. Rev. Lett.* **84**, 3807–3810 (2000).
14. I. Erez, E. R. Wallach, Y. Shagam, Simultaneous Enantiomer-Resolved Ramsey Spectroscopy Scheme for Chiral Molecules. *Phys. Rev. X* **13**, 41025 (2023).
15. O. Neufeld, O. Wengrowicz, O. Peleg, A. Rubio, O. Cohen, Detecting multiple chiral centers in chiral molecules with high harmonic generation. *Opt. Express* **30**, 3729–3740 (2022).
16. D. Habibović, K. R. Hamilton, O. Neufeld, L. Rego, Emerging tailored light sources for studying chirality and symmetry. *Nat. Rev. Phys.* (2024). https://doi.org/https://doi.org/10.1038/s42254-024-00763-8.
17. P. Fischer, D. S. Wiersma, R. Righini, B. Champagne, A. D. Buckingham, Three-Wave Mixing in Chiral Liquids. *Phys. Rev. Lett.* **85**, 4253–4256 (2000).
18. M. Leibscher, T. F. Giesen, C. P. Koch, Principles of enantio-selective excitation in three-wave mixing spectroscopy of chiral molecules. *J. Chem. Phys.* **151**, 14302 (2019).
19. D. Patterson, M. Schnell, J. M. Doyle, Enantiomer-specific detection of chiral molecules via microwave spectroscopy. *Nature* **497**, 475 (2013).
20. C. Pérez, *et al.*, Coherent Enantiomer-Selective Population Enrichment Using Tailored Microwave Fields. *Angew. Chemie Int. Ed.* **56**, 12512–12517 (2017).
21. S. Eibenberger, J. Doyle, D. Patterson, Enantiomer-Specific State Transfer of Chiral Molecules. *Phys. Rev. Lett.* **118**, 123002 (2017).
22. J. Lee, *et al.*, Quantitative Study of Enantiomer-Specific State Transfer. *Phys. Rev. Lett.* **128**, 173001 (2022).
23. D. Ayuso, *et al.*, Synthetic chiral light for efficient control of chiral light–matter interaction. *Nat. Photonics* **13**, 866–871 (2019).
24. O. Neufeld, *et al.*, Ultrasensitive Chiral Spectroscopy by Dynamical Symmetry Breaking in High Harmonic Generation. *Phys. Rev. X* **9**, 031002 (2019).
25. P. Fischer, F. W. Wise, A. C. Albrecht, Chiral and Achiral Contributions to Sum-Frequency Generation from Optically Active Solutions of Binaphthol. *J. Phys. Chem. A* **107**, 8232–8238 (2003).
26. J.-H. Choi, S. Cheon, M. Cho, Calculations of vibrationally resonant sum- and difference-frequency-generation spectra of chiral molecules in solutions: Three-wave-mixing vibrational optical activity. *J. Chem. Phys.* **132**, 74506 (2010).
27. M. A. Belkin, T. A. Kulakov, K.-H. Ernst, L. Yan, Y. R. Shen, Sum-Frequency Vibrational Spectroscopy on Chiral Liquids: A Novel Technique to Probe Molecular Chirality. *Phys. Rev. Lett.* **85**, 4474–4477 (2000).
28. M. A. Belkin, S. H. Han, X. Wei, Y. R. Shen, Sum-Frequency Generation in Chiral Liquids near Electronic Resonance. *Phys. Rev. Lett.* **87**, 113001 (2001).
29. G. J. Simpson, Molecular Origins of the Remarkable Chiral Sensitivity of Second-Order Nonlinear Optics. *ChemPhysChem* **5**, 1301–1310 (2004).
30. P. Fischer, Nonlinear Optical Spectroscopy of Chiral Molecules. *Chirality* **17**, 421–437 (2005).
31. D. Ayuso, A. F. Ordonez, P. Decleva, M. Ivanov, O. Smirnova, Enantio-sensitive unidirectional light bending. *Nat. Commun.* **12**, 3951 (2021).
32. D. Baykusheva, H. J. Wörner, Chiral Discrimination through Bielliptical High-Harmonic Spectroscopy. *Phys. Rev. X* **8**, 031060 (2018).
33. M. Pitzer, *et al.*, Direct Determination of Absolute Molecular Stereochemistry in Gas Phase by Coulomb Explosion Imaging. *Science* **341**, 1096–1100 (2013).
34. P. Herwig, *et al.*, Imaging the Absolute Configuration of a Chiral Epoxide in the Gas Phase. *Science* **342**, 1084–1086 (2013).
35. J.-L. Bégin, *et al.*, Nonlinear helical dichroism in chiral and achiral molecules. *Nat. Photonics* **17**, 82–88 (2023).
36. W. Brullot, M. K. Vanbel, T. Swusten, T. Verbiest, Resolving enantiomers using the optical angular momentum of twisted light. *Sci. Adv.* **2**, e1501349 (2023).
37. N. Mayer, *et al.*, Chiral topological light for detection of robust enantiosensitive observables. *Nat. Photonics* **18**, 1155–1160 (2024).
38. N. Böwering, *et al.*, Asymmetry in Photoelectron Emission from Chiral Molecules Induced by Circularly Polarized Light. *Phys. Rev. Lett.* **86**, 1187–1190 (2001).
39. M. H. M. Janssen, I. Powis, Detecting chirality in molecules by imaging photoelectron circular dichroism. *Phys. Chem. Chem. Phys.* **16**, 856–871 (2014).
40. S. B. and A. F. and R. G. and R. C. and D. D. and B. F. and N. F. and F. L. and S. P. and T. R. and V. B.

and Y. M. and B. Pons, Universality of photoelectron circular dichroism in the photoionization of chiral molecules. *New J. Phys.* **18**, 102002 (2016).
41. A. D. Müller, E. Kutscher, A. N. Artemyev, P. V Demekhin, Photoelectron circular dichroism in the multiphoton ionization by short laser pulses. III. Photoionization of fenchone in different regimes. *J. Chem. Phys.* **152**, 44302 (2020).
42. B. Ritchie, Theory of the angular distribution of photoelectrons ejected from optically active molecules and molecular negative ions. *Phys. Rev. A* **13**, 1411–1415 (1976).
43. I. Powis, Photoelectron circular dichroism of the randomly oriented chiral molecules glyceraldehyde and lactic acid. *J. Chem. Phys.* **112**, 301–310 (2000).
44. A. Comby, *et al.*, Relaxation Dynamics in Photoexcited Chiral Molecules Studied by Time-Resolved Photoelectron Circular Dichroism: Toward Chiral Femtochemistry. *J. Phys. Chem. Lett.* **7**, 4514–4519 (2016).
45. D. Faccialà, *et al.*, Time-Resolved Chiral X-Ray Photoelectron Spectroscopy with Transiently Enhanced Atomic Site Selectivity: A Free-Electron Laser Investigation of Electronically Excited Fenchone Enantiomers. *Phys. Rev. X* **13**, 11044 (2023).
46. P. V Demekhin, A. N. Artemyev, A. Kastner, T. Baumert, Photoelectron Circular Dichroism with Two Overlapping Laser Pulses of Carrier Frequencies ω and 2ω Linearly Polarized in Two Mutually Orthogonal Directions. *Phys. Rev. Lett.* **121**, 253201 (2018).
47. S. Rozen, *et al.*, Controlling Subcycle Optical Chirality in the Photoionization of Chiral Molecules. *Phys. Rev. X* **9**, 031004 (2019).
48. M. Hofmann, *et al.*, Sub-cycle resolved strong field ionization of chiral molecules and the origin of chiral photoelectron asymmetries. *Phys. Rev. Res.* **6**, 043176 (2024).
49. O. Neufeld, O. Cohen, Optical Chirality in Nonlinear Optics: Application to High Harmonic Generation. *Phys. Rev. Lett.* **120**, 133206 (2018).
50. O. Neufeld, M. Even Tzur, O. Cohen, Degree of chirality of electromagnetic fields and maximally chiral light. *Phys. Rev. A* **101**, 053831 (2020).
51. O. Neufeld, O. Cohen, Unambiguous definition of handedness for locally chiral light. *Phys. Rev. A* **105**, 23514 (2022).
52. O. Neufeld, H. Hübener, A. Rubio, U. De Giovannini, Strong chiral dichroism and enantiopurification in above-threshold ionization with locally chiral light. *Phys. Rev. Res.* **3**, L032006 (2021).
53. O. Neufeld, D. Podolsky, O. Cohen, Floquet group theory and its application to selection rules in harmonic generation. *Nat. Commun.* **10**, 405 (2019).
54. M. E. Tzur, O. Neufeld, E. Bordo, A. Fleischer, O. Cohen, Selection rules in symmetry-broken systems by symmetries in synthetic dimensions. *Nat. Commun.* **13**, 1312 (2022).
55. G. Lerner, *et al.*, Multiscale dynamical symmetries and selection rules in nonlinear optics. *Sci. Adv.* **9**, eade0953 (2023).
56. O. Neufeld, N. Tancogne-Dejean, U. De Giovannini, H. Hübener, A. Rubio, Light-Driven Extremely Nonlinear Bulk Photogalvanic Currents. *Phys. Rev. Lett.* **127**, 126601 (2021).
57. T. Weitz, *et al.*, Lightwave-driven electrons in a Floquet topological insulator. *arXiv Prepr. arXiv2407.17917* (2024).
58. M. Feldman, M. Even Tzur, O. Cohen, Selection Rules of Linear and Nonlinear Polarization-Selective Absorption in Optically Dressed Matter. *Photonics* (2024).
59. O. Neufeld, N. Tancogne-Dejean, U. De Giovannini, H. Hübener, A. Rubio, Attosecond magnetization dynamics in non-magnetic materials driven by intense femtosecond lasers. *npj Comput. Mater.* **9**, 39 (2023).
60. A. Fleischer, N. Moiseyev, Adiabatic theorem for non-Hermitian time-dependent open systems. *Phys. Rev. A* **72**, 032103 (2005).
61. S. Ito, *et al.*, Build-up and dephasing of Floquet–Bloch bands on subcycle timescales. *Nature* (2023). https://doi.org/10.1038/s41586-023-05850-x.
62. M. A. L. Marques, *et al.*, "Time-Dependent Density Functional Theory" in *Time-Dependent Density Functional Theory*, (Springer, 2003).
63. A. Castro, *et al.*, octopus: a tool for the application of time-dependent density functional theory. *Phys. status solidi* **243**, 2465–2488 (2006).
64. X. Andrade, *et al.*, Real-space grids and the Octopus code as tools for the development of new simulation approaches for electronic systems. *Phys. Chem. Chem. Phys.* **17**, 31371–31396 (2015).
65. N. Tancogne-Dejean, *et al.*, Octopus, a computational framework for exploring light-driven phenomena and quantum dynamics in extended and finite systems. *J. Chem. Phys.* **152**, 124119 (2020).
66. C. Hartwigsen, S. Goedecker, J. Hutter, Relativistic separable dual-space Gaussian pseudopotentials from H to Rn. *Phys. Rev. B* **58**, 3641–3662 (1998).
67. C. Legrand, E. Suraud, P.-G. Reinhard, Comparison of self-interaction-corrections for metal clusters. *J.*

*Phys. B At. Mol. Opt. Phys.* **35**, 1115–1128 (2002).
68. L. Tao, A. Scrinzi, Photo-electron momentum spectra from minimal volumes: The time-dependent surface flux method. *New J. Phys.* **14**, 013021 (2012).
69. A. Scrinzi, t-SURFF: fully differential two-electron photo-emission spectra. *New J. Phys.* **14**, 085008 (2012).
70. P. Wopperer, U. De Giovannini, A. Rubio, Efficient and accurate modeling of electron photoemission in nanostructures with TDDFT. *Eur. Phys. J. B* **90**, 51 (2017).
71. S. A. Sato, H. Hübener, A. Rubio, U. De Giovannini, First-principles simulations for attosecond photoelectron spectroscopy based on time-dependent density functional theory. *Eur. Phys. J. B* **91**, 126 (2018).
72. U. De Giovannini, H. Hübener, A. Rubio, A First-Principles Time-Dependent Density Functional Theory Framework for Spin and Time-Resolved Angular-Resolved Photoelectron Spectroscopy in Periodic Systems. *J. Chem. Theory Comput.* **13**, 265–273 (2017).
73. V. Averbukh, O. E. Alon, N. Moiseyev, High-order harmonic generation by molecules of discrete rotational symmetry interacting with circularly polarized laser field. *Phys. Rev. A* **64**, 033411 (2001).
74. O. E. Alon, V. Averbukh, N. Moiseyev, Atoms, Molecules, Crystals and Nanotubes in Laser Fields: From Dynamical Symmetry to Selective High-Order Harmonic Generation of Soft X-Rays. *Adv. Quantum Chem.* **47**, 393–421 (2004).
75. O. E. Alon, V. Averbukh, N. Moiseyev, Selection Rules for the High Harmonic Generation Spectra. *Phys. Rev. Lett.* **80**, 3743 (1998).
76. D. Shafir, Y. Mairesse, D. M. Villeneuve, P. B. Corkum, N. Dudovich, Atomic wavefunctions probed through strong-field light–matter interaction. *Nat. Phys.* **5**, 412–416 (2009).
77. L. Zhang, *et al.*, Laser-sub-cycle two-dimensional electron-momentum mapping using orthogonal two-color fields. *Phys. Rev. A* **90**, 61401 (2014).
78. S. Stremoukhov, *et al.*, Origin of ellipticity of high-order harmonics generated by a two-color laser field in the cross-polarized configuration. *Phys. Rev. A* **94**, 013855 (2016).
79. D. Habibović, W. Becker, D. B. Milošević, Symmetries and Selection Rules of the Spectra of Photoelectrons and High-Order Harmonics Generated by Field-Driven Atoms and Molecules. *Symmetry (Basel).* (2021).
80. D. B. Milošević, W. Becker, R. Kopold, S. W., High-harmonic generation by a bichromatic bicircular laser field. *Laser Phys.* **11**, 165–168 (2001).
81. A. Fleischer, O. Kfir, T. Diskin, P. Sidorenko, O. Cohen, Spin angular momentum and tunable polarization in high-harmonic generation. *Nat. Photonics* **8**, 543–549 (2014).
82. O. Kfir, *et al.*, Generation of bright phase-matched circularly-polarized extreme ultraviolet high harmonics. *Nat. Photonics* **9**, 99–105 (2015).
83. X.-M. Tong, S.-I. Chu, Generation of circularly polarized multiple high-order harmonic emission from two-color crossed laser beams. *Phys. Rev. A* **58**, 2656(R) (1998).
84. O. Neufeld, O. Cohen, Background-Free Measurement of Ring Currents by Symmetry-Breaking High-Harmonic Spectroscopy. *Phys. Rev. Lett.* **123**, 103202 (2019).